\documentstyle[aps,floats,epsf,preprint]{revtex}

\begin{document}
\title
{Axial vector tetraquak with $S=+2$}

\author{Y. Kanada-En'yo, O. Morimatsu, and T. Nishikawa}

\address{Institute of Particle and Nuclear Studies, \\
High Energy Accelerator Research Organization,\\
1-1 Oho, Tsukuba, Ibaraki 305-0801, Japan}

\maketitle

\begin{abstract}
The possibility of an axial vector isoscalar 
tetraquark with $ud\bar{s}\bar{s}$ 
is discussed. 
If the pentaquark $\Theta^+(1540)$ has the 
$(qq)_{\bar{3}}(qq)_{\bar{3}}\bar{q}$ configuration,
the isoscalar $ud\bar{s}\bar{s}$($\vartheta^+$-meson) state with $J^P=1^+$
is expected to exist in the mass region 
lower than, or close to, the mass of $\Theta^+(1540)$. 
Within a flux-tube quark model, 
a possible resonant state of 
$ud\bar{s}\bar{s}(J^{P}=1^{+})$ is suggested to appear at around 1.4 GeV 
with the width ${\cal{O}}(20\sim 80)$ MeV.
We propose that the $\vartheta^+$-meson is a good candidate for the 
tetraquark search, which would be observed in the $K^+K^+\pi^-$ decay channel.
\end{abstract}


\section{Introduction}

The possibility of multiquark states has been discussed for a long time
\cite{Jaffe-4q,Jaffe-h,Jaffe79,Isgur-4q,close87,carlsonb,lipkin97,Stancu98,Sakai,Barnes,Oka-rev}.
In particular, the possible $qq\bar{q}\bar{q}$ states have been
suggested in many theoretical efforts to understand light scalar 
mesons (see, for example, Refs.\cite{Jaffe-4q,Jaffe79,Isgur-4q}). 
The $4q$ states 
were proposed in descriptions of $f_0(600)$ and $f_0(980)$, 
where the strong attraction between 
$(qq)_{\bar{3}}$ and $(\bar{q}\bar{q})_{3}$ plays an important role 
\cite{Jaffe-4q,Jaffe79}. Here, $(qq)_{\bar{3}}$ and $(\bar{q}\bar{q})_{3}$
denote the color-anti-triplet quark pair and the color-triplet anti-quark
pair, respectively.
On the other hand, the $KK$ molecule 
states were suggested to understand the properties of $f_0(980)$ and
$a_0(980)$\cite{Isgur-4q}. 
Since the masses of negative-parity mesons are expected to be above 
1 GeV in a naive interpretation with $P$-wave $q\bar{q}$ states, 
it is considered that these scalar mesons below 1 GeV may be
hybrids of $P$-wave $q\bar{q}$ and 
compact $(qq)_{\bar{3}}(\bar{q}\bar{q})_3$ 
with meson-meson tails in the outer region, as argued in Ref.\cite{close02}. 
Even if the $4q$ components are dominant in a certain
meson whose minimal content is $2q$,
it is difficult to find direct evidence of the $4q$ components 
due to mixing with the conventional $q\bar{q}$ state 
via the annihilation of $q\bar{q}$ pairs.
Our main interest here is the possibility of narrow ``tetraquark'' states, 
whose minimal quark content is 4$q$.

The recent observation of $D_{sJ}$(2317)\cite{Ds} and 
reports of the pentaquark baryon $\Theta^+(uudd\bar{s})$
\cite{leps,clasb,diana,clasa,saphir,itep,hermes,itep-2,zeus} 
revived motivation for 
experimental and theoretical studies on multiquarks in hadron physics,
though the existence of $\Theta^+$ is yet to be well established.
One of the striking characteristics of the $\Theta^+$ is its narrow
width. For a theoretical interpretation of
why the $\Theta^+$ is extremely narrow, 
the possibility of the spin-parity $J^P=1/2^+$
and $J^P=3/2^-$ has been discussed by many groups 
\cite{Oka-rev,jaffe,karliner,hosaka,ENYO-penta,takeuchi,eidemuler,nishikawa}. 
Since only the
$P$-wave and $D$-wave are allowed in $NK$ decays from 
the $J^P=1/2^+$ and $J^P=3/2^-$ states, respectively, 
the width should be suppressed due to a high centrifugal barrier.
The transition into meson-baryon states should be further suppressed if
the pentaquark has an exotic color configuration 
$(qq)_{\bar{3}}(qq)_{\bar{3}}\bar{q}$. The factor $1/3$ in the transition 
appears from the overlap of the color wave functions of quarks.
In addition to suppression due to the color degrees of freedom of 
5 quarks, another suppression effect can be considered in 
flux-tube pictures\cite{carlsonb,ENYO-penta,song,bando,suganuma}, because 
the transitions between different flux-tube 
topologies are suppressed due to a rearrangement of the gluon field.
This means that the decays from such 
exotic flux-tube configurations as $(qq)_{\bar{3}}(qq)_{\bar{3}}\bar{q}$
(Fig.~ \ref{fig:flux-4q}(e)) 
into meson-baryon-like $(qqq)_{1}(q\bar{q})_{1}$ 
(Fig.~ \ref{fig:flux-4q}(f)) are suppressed. 
In general, the different flux-tube configuration
appears in multiquarks that contain
more than 3 quarks, as shown in Fig.~ \ref{fig:flux-4q}, and the coupling 
between the disconnected tube and the connected tube topologies should be 
strongly suppressed. 

The predicted spin-parity $J^P=1/2^+$ and $J^P=3/2^-$ of 
the pentaquark $uudd\bar{s}$ are abnormal in a naive quark picture, 
where the $J^P=1/2^-$ should be the lowest state, 
while other spin-parity states
are expected to be highly excited.
Originally, a narrow $J^P=1/2^+$ was predicted 
in the Skyrme soliton model by Diakonov et al.\cite{diakonov}. 
For a theoretical explanation of the $J^P=1/2^+$ state from 
the point of view of the constituent quarks, the diquark picture
and the triquark picture are proposed in Refs.\cite{jaffe,karliner}.
Recently, constituent quark model calculations \cite{ENYO-penta,takeuchi}  
suggested an abnormal level structure in 
the compact $uudd\bar{s}$ system with the
$(qq)_{\bar{3}}(qq)_{\bar{3}}\bar{q}$ configuration,
where the masses of the $J^P=1/2^+$ 
and/or $J^P=3/2^-$ states may degenerate with 
the $J^P=1/2^-$ state. 

We now turn to a discussion on the possibility of tetraquarks.
By replacing a $ud$-diquark in the $\Theta^+$ with an $\bar{s}$ quark, 
it is natural to expect that a tetraquark with the $ud\bar{s}\bar{s}$ content
may exist at nearly the same energy region. 
In order to search for a narrow tetraquark, we follow the analogy of 
the theoretical explanation why the pentaquark $\Theta^+$ can be narrow.  
Firstly, one should consider those states with
unnatural spin and parity, which cannot decay into two light hadrons
(psuedscalar mesons) in the $S$-wave channel. 
Second, the exotic flux-tube configurations with connected tubes
(Fig.~ \ref{fig:flux-4q}(c) and (e)) would be 
essential to stabilize the exotic hadrons.
We then propose a $J^P=1^+$ $ud\bar{s}\bar{s}$ state 
with the $(qq)_{\bar{3}}(\bar{q}\bar{q})_{3}$ configuration 
as a candidate of narrow tetraquark states. 
It should be stressed that two-body $KK$ decays from any 
$J^P=1^+$ $ud\bar{s}\bar{s}$ state are forbidden because of 
conservation of the total spin and parity. The lowest threshold energy 
of the allowed two-body decays is 1.39 GeV for the $KK^*(895)$ channel.
If the mass of the $J^P=1^+$ $ud\bar{s}\bar{s}$ state
lies below the $KK^*$, two-meson decay channels are closed, and hence
its width must be narrow. 

The 4q states have been proposed by Jaffe in 1977
\cite{Jaffe-4q}.
The $(qq)_{\bar 3}$ diquark and 
the $(\bar{q}\bar{q})_{3}$ anti-diquark correlations 
play an important role in the stability of the $4q$ state,
because there exists attraction between $(qq)_{\bar 3}$
and $(\bar{q}\bar{q})_{3}$ due to 
the confining and the one-gluon-exchange(OGE) potential.
Moreover, it is known that the spin-zero flavor-singlet $(qq)_{\bar{3}}$ 
diquark is favored, 
because it gains the color-magnetic interaction. 
Thus, the diquark(anti-diquark) correlation leads to 
confined $4q$ states with the $(qq)_{\bar{3}}(\bar{q}\bar{q})_{3}$
color configuration, which might make the $4q$ system compact and stable.
In the $J^P=1^+$ state of the tetraquark 
$(ud)_{\bar{3}}(\bar{s}\bar{s})_{3}$, one can naturally 
expect that the lowest is the isoscalar 
$J^P=1^+$ state with a spin-zero $(ud)_{\bar{3}}$ 
and a spin-one $(\bar{s}\bar{s})_3$ in a spatially symmetric orbit.
In the spatially symmetric $(\bar{s}\bar{s})_{3}$,
the spin-zero(spin-singlet) configuration is forbidden, and hence 
the $(\bar{s}\bar{s})_{3}$ must have spin-one.
Although the spin-one $(\bar{s}\bar{s})_{3}$ feels some
repulsive color-magnetic interaction, 
the repulsion is expected to be small because the color-magnetic term 
in the OGE potential is suppressed by the quark-mass factor $m^{-2}$.
In the flux-tube model, the $(qq)_{\bar{3}}(\bar{q}\bar{q})_3$ state
has the exotic tube topology shown in Fig.~ \ref{fig:flux-4q}(c). 
Therefore, its coupling with  
two-hadron states should be small, due to the suppressed 
transitions between the different tube topologies, (c) and (d).  


The tetraquark $ud\bar{s}\bar{s}$ states are
discussed in Ref.\cite{Jaffe-4q}, and noted as $E_{(KK)}$-mesons.
In Ref.\cite{Jaffe-4q}, the theoretical mass for the 
isoscalar $ud\bar{s}\bar{s}(J^P=1^+)$ 
state is predicted to be 1.65 GeV in the MIT bag model\cite{Jaffe-4q}.
Recently, the isoscalar $ud\bar{s}\bar{s}$
in the flavor $\bar{10}$ group
was suggested in analogy with the $\Theta^+$ by T. Burns et al.\cite{Close-4q}
and by Karliner and Lipkin \cite{karliner04}.
In Ref.\cite{karliner04}, the possibility of a $0^+$ state is discussed. 
However, we should remark that the $J^P=0^+$ is not allowed 
in the isoscalor $ud\bar{s}\bar{s}$ system
within the spatially symmetric configuration, and hence, 
the $J^P=0^+$ is expected to be unfavored.
The tetraquark $ud\bar{s}\bar{s}$ is called a $\vartheta^+$-meson 
in Ref.\cite{Close-4q}, where the $J^P=1^-$ state with 
the orbital angular momentum $L=1$ is 
predicted in the mass region $\sim$ 1.6 GeV.
Although the $J^P=1^-$ state may gain color-magnetic attraction,
it needs the $L=1$ excitation energy, and is 
expected to be higher than the spatially symmetric state.
Another claim for the $\vartheta^+(J^P=1^-)$ state is that it 
can decay into $P$-wave $KK$ states. The centrifugal barrier
may not be high enough to stabilize the state much above the threshold energy. 
Therefore, we think that the 
$\vartheta^+$($J^P=1^+$) is a better candidate for 
narrow tetraquarks.

In this paper, we consider the $\vartheta^+$-meson with $J^P=1^+$ 
by a constituent quark model.
The theoretical method of the calculations is the same as that applied
to the pentaquark study in Ref.\cite{ENYO-penta}.
Namely, we apply the flux-tube quark model with antisymmetrized molecular
dynamics(AMD)\cite{ENYObc,AMDrev} to the $4q$ systems.
Based on the picture of a flux-tube model, we ignore the coupling 
between configurations shown in Fig.~ \ref{fig:flux-4q}(c) and 
Fig.~ \ref{fig:flux-4q}(d), and solve 
the $4q$ dynamics with the variational method in the model space 
$(qq)_{\bar{3}}(\bar{q}\bar{q})_3$ shown in
Fig.~ \ref{fig:flux-4q}(c). 
The Coulomb and color-magnetic terms of the OGE potential 
and the string potential are taken into account.
In order to evaluate the $\vartheta^+(J^P=1^+)$ mass, 
we adopt the observed $\Theta^+$ mass as an input 
as well as the normal hadron spectra.
We also try to interpret a $f_1$-meson in $1.4\sim 1.6$ GeV region with 
the 4 $q$ state, which would help to check
the reliability of the present calculations.
The widths of these states are also discussed.

\section{Hamiltonian}

The adopted Hamiltonian is the same as that of previous work\cite{ENYO-penta}
as $H=H_0+H_I+H_f$, where $H_0$ consists of the mass and kinetic terms,
$H_I$ represents the short-range OGE interaction, and $H_f$ is the 
string potential given by the energy of the flux tubes.
The quarks are treated as non-relativistic spin-$\frac{1}{2}$ Fermions.
The OGE potential consists of Coulomb and the color-magnetic interactions,
as
\begin{equation}
H_I=\alpha_c\sum_{i<j}F_i F_j
[\frac{1}{r_{ij}}-\frac{2\pi}{3m_i m_j}s(r_{ij})
\sigma_i\cdot\sigma_j]. 
\end{equation}
Here, $\alpha_c$ is the quark-gluon coupling constant, and 
$F_iF_j$ is defined by 
$\sum_{\alpha=1,\cdots,8} F^\alpha_i F^\alpha_j$,
where $F^\alpha_i$ is the 
generator of color $SU(3)$, $\frac{1}{2}\lambda^\alpha_i$ for quarks and
$-\frac{1}{2}(\lambda^\alpha_i)^*$ for anti-quarks. $m_i$ is the quark mass
$m_q$ for $u$ and $d$ quarks, and $m_s$ for a $s$ quark. 
The usual $\delta(r_{ij})$ function in the spin-spin 
interaction is replaced by
a finite-range Gaussian, $s(r_{ij})
\equiv\left[ \frac{1}{2\sqrt{\pi}\Lambda}\right]^3\exp{
\left[-\frac{r_{ij}^2}{4\Lambda^2}\right]}$.

In the flux-tube quark model \cite{carlsonb,carlson},
the confining string potential is written as
$H_{f}={\sigma}L_f-M^0$,
where ${\sigma}$ is the string tension, $L_f$ is
the minimum length of the flux tubes, and $M^0$ is the 
zero-point string energy. 
For the meson and 3$q$-baryon systems, the flux tube configurations 
are the linear line and the $Y$-type configuration with a junction,
as shown in Figs.~ \ref{fig:flux-4q}(a) and (b), respectively. 
For the $q^2\bar{q}^2$ mesons and $q^4\bar{q}$ baryons,
the exotic topologies Figs.~ \ref{fig:flux-4q}(c) and (e) appear corresponding to 
the $(qq)_{\bar{3}}(\bar{q}\bar{q})_3$ and
$(qq)_{\bar{3}}(qq)_{\bar{3}}\bar{q}$,
in addition to the normal two-hadron configurations
(Fig.~ \ref{fig:flux-4q}(d) and (f)).
In principle, besides these color configurations, 
other color configurations are possible in totally 
color-singlet $q^2\bar{q}^2$ and $q^4\bar{q}$ systems 
by incorporating a color-symmetric $(qq)_6$ pair.
However, since such a string from the $(qq)_6$ is an excited one, and is 
unfavored in the strong-coupling limit of lattice QCD\cite{Kogut75},
we should consider only color-$3$ flux tubes 
as the elementary tubes.
The string potentials given by the tube lengths of the configuration
Fig.~ \ref{fig:flux-4q}(b), (c) and (e) are supported by lattice QCD calculations
\cite{Takahashi,okiharu-4q,okiharu-5q}.

In the practical calculation of the expectation values of the string potential
$\langle\Phi|H_f|\Phi\rangle$ 
with respect to a meson state($\Phi_{q\bar{q}}$), 
a three-quark state($\Phi_{q^3}$), a $(qq)_{\bar{3}}(\bar{q}\bar{q})_3$
state($\Phi_{(qq)_{\bar{3}}(\bar{q}\bar{q})_3}$), and a 
$(qq)_{\bar{3}}(qq)_{\bar{3}}\bar{q}$ state
($\Phi_{(qq)_{\bar{3}}(qq)_{\bar{3}}\bar{q}}$),
the minimum length of the flux tubes $L_f$ is approximated by
a linear combination of two-body distances $r_{ij}$ as, 
\begin{eqnarray}
\label{eq:2q}
L_f&\approx& r_{1\bar{1}} 
\quad {\rm in \ }\langle\Phi_{q\bar{q}}|H_f|\Phi_{q\bar{q}}\rangle,\\
\label{eq:3q}
L_f&\approx& \frac{1}{2}(r_{12}+r_{23}+r_{31})
\quad {\rm in \ }\langle\Phi_{q^3}|H_f|\Phi_{q^3}\rangle,\\
\label{eq:4q}
L_f&\approx& \frac{1}{2}(r_{12}+r_{\bar{1}\bar{2}})+
\frac{1}{4}(r_{1\bar{1}}+r_{1\bar{2}}+r_{2\bar{1}}+r_{2\bar{2}})
\quad {\rm in \ }\langle\Phi_{(qq)_{\bar{3}}(\bar{q}\bar{q})_3}|H_f|
\Phi_{(qq)_{\bar{3}}(\bar{q}\bar{q})_3} \rangle,\\
L_f&\approx& \frac{1}{2}(r_{12}+r_{34})+
\frac{1}{8}(r_{13}+r_{14}+r_{23}+r_{24})
+\frac{1}{4}(r_{\bar{1}1}+r_{\bar{1}2}+r_{\bar{1}3}+r_{\bar{1}4})\nonumber\\
\label{eq:5q}
&& \quad {\rm in \ }
\langle\Phi_{(qq)_{\bar{3}}(qq)_{\bar{3}}\bar{q}}|H_f|
\Phi_{(qq)_{\bar{3}}(qq)_{\bar{3}}\bar{q}}\rangle.
\end{eqnarray}

$M^0$ depends on the flux-tube topology and is denoted here as 
$M^0_{q\bar{q}}$, $M^0_{q^3}$, $M^0_{[qq][\bar{q}\bar{q}]}$ and
$M^0_{[qq][qq]\bar{q}}$ for the configurations shown in 
Fig.~ \ref{fig:flux-4q}(a), (b), (c), and (e), respectively. 
($[qq]$ and $[\bar{q}\bar{q}]$ indicate $(qq)_{\bar{3}}$
and $(\bar{q}\bar{q})_3$, respectively.)

In the present calculation, we ignore other terms such as 
tensor and spin-orbit interactions in the OGE potential, and we do not
introduce flavor-exchange interactions. As shown later, 
the major properties of the normal hadron mass spectra is qualitatively 
reproduced by the present Hamiltonian.

\section{Model wave function and parameters}

We solve the eigen states of the Hamiltonian with a variational method
in the AMD model space proposed in the previous paper.
The AMD wave function in a quark model is given as follows. 
\begin{eqnarray}
&\Phi({\bf Z})=(1\pm P) {\cal A}
\left[\phi_{Z_1}\phi_{Z_2}\cdots\phi_{Z_{N_q}} \Phi^S\Phi^X \right],\\
&\phi_{Z_i}=\left(\frac{1}{\pi b^2}\right)^{3/4}
\exp\left[-\frac{1}{2b^2}(r-\sqrt{2}bZ_i)^2+\frac{1}{2}Z^2_i\right],
\end{eqnarray}
where $1\pm P$ is the parity projection operator, ${\cal A}$ is the
anti-symmetrization operator,  
and the spatial part $\phi_{Z_i}$ of 
the $i$th single-particle wave function
given by a Gaussian whose center is located at $Z_i$ in phase space. 
The spin function, $\Phi^S$, is given as
\begin{equation}
\Phi^S=\sum_{m_{1},\cdots,m_{N_q}} 
c_{m_{1}\cdots m_{N_q}}
|m_{1}\cdots m_{N_q}\rangle_S,
\end{equation}
where $|m\rangle_S (m=\uparrow,\downarrow)$ is the intrinsic-spin function. 
$\Phi^X$ is the flavor-color function.
For example, the flavor-color function for 
the tetraquark $ud\bar{s}\bar{s}$ system with color-configuration
$(qq)_{\bar{3}}(\bar{q}\bar{q})_3$ is written as
\begin{equation}
\Phi^X=|ud\bar{s}\bar{s}\rangle\otimes  
\epsilon_{abc}\epsilon_{efc}|ab\bar{e}\bar{f}\rangle_C.
\end{equation}
In the present wave function 
we do not explicitly perform isospin projection, but
the wave functions obtained by energy variation 
are found to be approximately isospin-eigen states
in most cases due to the color-spin symmetry.

\begin{figure}
\noindent
\epsfxsize=0.35\textwidth
\centerline{\epsffile{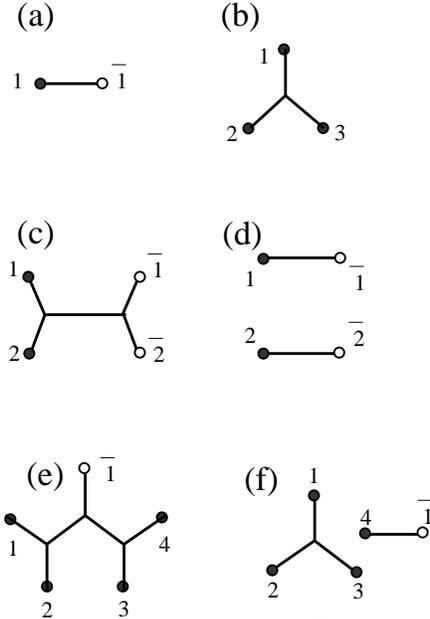}}
\caption{\label{fig:flux-4q}
Flux-tube configurations for $q\bar{q}$ meson (a),
$q^3$ baryon (b), $q^2\bar{q}^2$ states (c),(d), and $q^4\bar{q}$ states
(e),(f). 
For the $q^2\bar{q}^2$ states, the exotic tube configuration (c) 
corresponds to the $(qq)_{\bar{3}}(\bar{q}\bar{q})_{3}$, and the 
disconnected tube (d) represents the meson-meson state,
$(q\bar{q})_{1}(q\bar{q})_{1}$. The configurations, 
$(qq)_{\bar{3}}(qq)_{\bar{3}}\bar{q}$ and $(qqq)_1(q\bar{q})_1$
for the $q^4\bar{q}$ system are illustrated in figures, (e) and (f), 
respectively. }
\end{figure}

As already mentioned, different flux-tube topologies appear in 
each of the $q^2\bar{q}^2$ and the $q^4\bar{q}$ systems.
Since the transitions between the different string configurations 
are of higher order in the strong coupling expansion, 
we ignore the coupling and perform a variational calculation
within a single flux-tube topology. 
In the present calculations, we adopt 
only the connected flux-tube configurations given in 
Figs. \ref{fig:flux-4q}(c) 
and (e), because we are interested in the confined and narrow states.
This is regarded as a kind of bound state approximations.

In the numerical calculation, the linear and Coulomb potentials
are approximated by seven-range Gaussians.
We use the same parameters as those adopted in Ref.\cite{ENYO-penta}:
\begin{eqnarray}
& \alpha_c =1.05, \nonumber\\
& \Lambda =0.13\ {\rm fm}, \nonumber\\
& m_q =0.313\ {\rm GeV}, \nonumber\\
& m_s =0.513\ {\rm GeV}, \nonumber\\
& {\sigma} =0.853\ {\rm GeV/fm}. 
\end{eqnarray}
Here, the quark-gluon coupling constant ($\alpha_c$) and the string
tension ($\sigma$) are chosen so as to fit the mass splitting among 
$N$, $\Delta$ and $N^*(1520)$. 
The width parameter ($b$) is chosen to be $0.5$ fm.

\begin{figure}
\noindent
\centerline{\epsfxsize=0.8\textwidth\epsffile{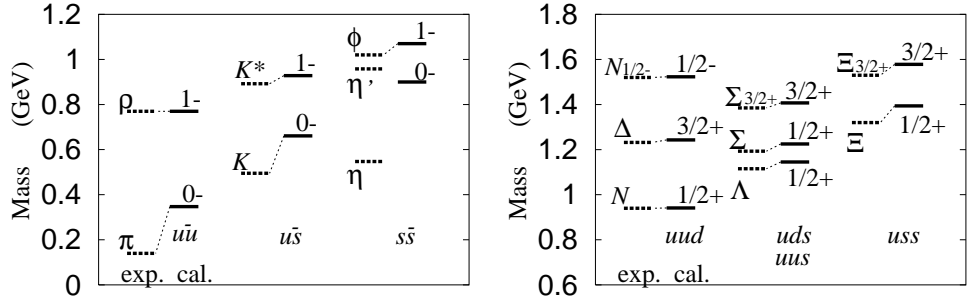}}
\caption{\label{fig:2q3q}
Mass spectra for $q\bar{q}$ mesons and $3q$ baryons. 
The experimental and calculated masses are shown by dashed and solid lines,
respectively.
We adjust the zero-point energy of the string potential for the $q\bar{q}$ 
system as $M^0_{q\bar{q}}=584 $ MeV to fit the experimental $\rho$-meson
mass. For the $3q$ system, $M^0_{q^3}$ is chosen to be
$M^0_{q^3}= 972$ MeV to reproduce the nucleon masses.}
\end{figure}

\section{Results and discussions}
\subsection{mesons and baryons}

In Fig.~ \ref{fig:2q3q}, we display the calculated masses of the conventional
mesons and baryons compared with the experimental data.
The zero-point energy of the string potential for the 3q system 
is chosen to be 
$M^0_{q^3}=972$ MeV to fit the nucleon mass, while 
$M^0_{q\bar{q}}$ for the $q\bar{q}$ is adjusted to be 584 MeV 
to reproduce the $\rho$-meson mass.
It is shown that 
the systematics of the mass spectra are reasonably reproduced by the
present calculations, except for the pseudoscalar mesons.

\subsection{$\vartheta^+$-meson($I=0$, $J^P=1^+$)}

As mentioned before, the zero-point energy ($M^0$) of the string potential 
depends on the flux-tube topology. We, here, phenomenologically 
deduce the unknown $M_{[qq][\bar{q}\bar{q}]}$ for the $4q$ system
with the help of the systematics of $M_{q\bar{q}}$, $M_{q^3}$ and
$M_{[qq][qq]\bar{q}}$ for the normal meson, baryon and pentaquark systems.
In a previous paper, we applied the present method 
to the $uudd\bar{s}$ system and studied the
properties of $\Theta^+$. In the results, it was predicted that 
the three narrow states, $I=0, J^P=1/2^+,3/2^+$ states and the
$I=1, J^P=3/2^-$ state may degenerate in almost the same mass region.
$M_{[qq][qq]\bar{q}}=2375$ MeV was chosen to fit the theoretical mass 
to the observed $\Theta^+$ mass. 

\begin{figure}
\noindent
\centerline{{\epsfxsize=0.45\textwidth\epsffile{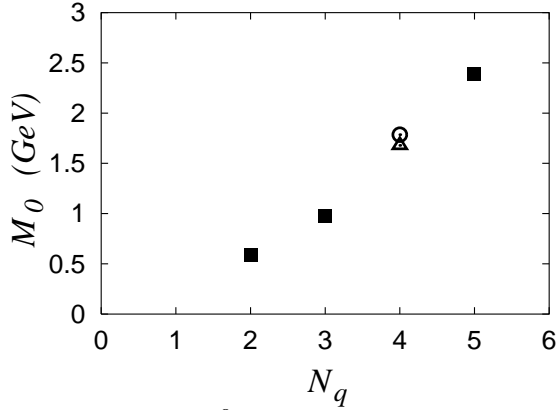}}}
\caption{\label{fig:m0}
Adopted zero-point energy ($M^0$) of the string potential as a
function of the quark number $N_q$.
$M^0_{q\bar{q}}$, $M^0_{q^3}$ and 
$M^0_{[qq][qq]\bar{q}}$ are adjusted to reproduce 
the $\rho$-meson, nucleon and
$\Theta^+$ masses, respectively.
The $M^0_{[qq][\bar{q}\bar{q}]}$ for the $(qq)_{\bar{3}}(\bar{q}\bar{q})_3$ 
is deduced by assuming the linear function $M^0(N_q)=a_0 N_q+b_0$,
where the parameters $a_0$ and $b_0$ are determined by fitting 
(i) 
$M^0(N_q=2)=M^0_{q\bar{q}}=584$ MeV and
 $M^0(N_q=5)=M_{[qq][qq]\bar{q}}=2375$ MeV, or 
(ii) $M^0(N_q=3)=M^0_{q^3}=972$ MeV and
$M^0(N_q=5)=M_{[qq][qq]\bar{q}}=2375$ MeV.
The circle and triangle indicate the $M^0_{[qq][\bar{q}\bar{q}]}$ values
obtained by the former fitting (i) 
and the latter one (ii), respectively. 
}
\end{figure}

In Fig.~ \ref{fig:m0}, the adopted $M_{q\bar{q}}$, $M_{q^3}$ and
$M_{[qq][qq]\bar{q}}$ are shown as a function of the quark number
($N_q$).
If the string potential is assumed to be 
a two-body linear-potential, $-a_s\sum_{ij}F_iF_j(r-r_0)$,
the potential is equivalent to the approximated flux-tube potential
in Eqs.~ \ref{eq:2q}-\ref{eq:5q} 
with the relations $\sigma=\frac{4}{3}a_s$ and  
$M^0=\frac{\sigma r_0}{2}N_q$. 
As a result, $M^0$ should be proportional to $N_q$ in the pairwise confining
potential, which leads to the relation 
$M_{[qq][qq]\bar{q}}/M_{q^3}=5/3$.
However, as shown in Fig.~ \ref{fig:m0}, the ratio  
$M_{[qq][qq]\bar{q}}/M_{q^3}$ is larger than $5/3$ 
in the present model, and also in the pentaquark study with 
a constituent quark-model calculation in Ref.\cite{takeuchi}.
This means that we need an extra attraction in the $[qq][qq]\bar{q}$ system 
in addition to the pair-wise confining potential
to understand the absolute mass of the 
$\Theta^+$(1.54) within constituent quark models.
We consider the dependence of $M^0$ on the tube topology
as the many-body potential, and we here accept the 
value $M_{[qq][qq]\bar{q}}=2375$ MeV \cite{ENYO-penta}
adjusted to the experimental $\Theta^+$ mass.
Then we phenomenologically determine the $M^0_{[qq][\bar{q}\bar{q}]}$ for 
$4q$ states by using the $M^0_{[qq][qq]\bar{q}}$ as an input as follows.

We assume the linear function 
$M^0(N_q)=a_0 N_q+b_0$ and determine the parameters
$(a_0,b_0)$ by fitting the $M_0$ values for the $q\bar{q}$ and
$[qq][qq]\bar{q}$ systems as 
(i) $M^0(N_q=2)=M^0_{q\bar{q}}=584$ MeV and
 $M^0(N_q=5)=M_{[qq][qq]\bar{q}}=2375$ MeV.
We also use another parameter set for $(a_0,b_0)$
by fitting the $M_0$ values for the $q^3$ and
$[qq][qq]\bar{q}$ systems as 
(ii) $M^0(N_q=3)=M^0_{q^3}=972$ MeV and
$M^0(N_q=5)=M_{[qq][qq]\bar{q}}=2375$ MeV.
With the obtained parameter sets $a_0$ and $b_0$, 
we obtain $M_{[qq][\bar{q}\bar{q}]}=1785$ MeV and
$M_{[qq][\bar{q}\bar{q}]}=1679$ MeV from $M^0(N_q)=a_0 N_q+b_0$, ($N_q=4$)
for the former fitting (i) and the latter one (ii), respectively.

\begin{figure}
\noindent
\centerline{\epsfxsize=0.8\textwidth\epsffile{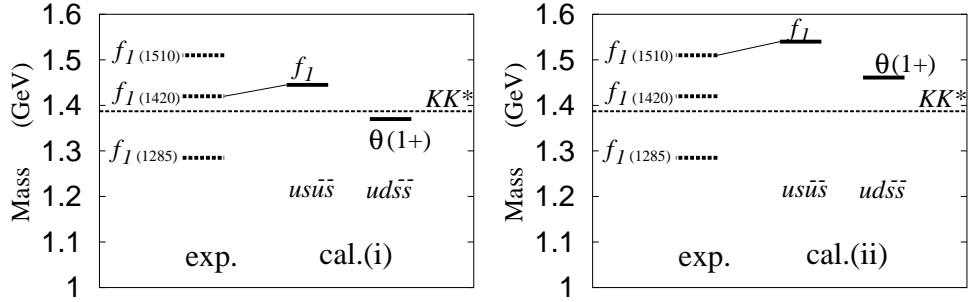}}
\caption{\label{fig:4q}
Masses of the $\vartheta^+(J^{PC}=1^+)$-meson and
$f_1$-mesons. The theoretical values (solid lines) shown in the left and
right panels were obtained 
by using (i) $M^0_{[qq][\bar{q}\bar{q}]}=1785$ MeV and
(ii) $M^0_{[qq][\bar{q}\bar{q}]}=1679$ MeV, respectively.
The mass of the $f_1$ with the $4q$ component was obtained by calculations
of the $us\bar{u}\bar{s}$ with $J^{PC}=1^{++}$.
The dashed lines are experimental masses of the $f_1$-mesons.
The experimental $KK^*$ threshold energy is also displayed by dotted lines.}
\end{figure}

Now, we apply the flux-tube quark model with AMD to the 
$(ud)_{\bar{3}}(\bar{s}\bar{s})_3$ system and 
calculate the $\vartheta^+(J^P=1^+)$ mass.
We use the above-determined zero-point energies,
 (i)$M^0_{[qq][\bar{q}\bar{q}]}$=1785 MeV 
and (ii)$M^0_{[qq][\bar{q}\bar{q}]}$=1679 MeV. 
The calculated $\vartheta^+$($I=0$,$J^P=1^+$) mass is 1.37 GeV in case~ (i)
and 1.46 GeV in case~ (ii).
The results indicate that the $\vartheta^+(J^P=1^+)$-meson 
may exist around 1.4 GeV, near the $KK^*$ threshold
(Fig.~ \ref{fig:4q}).

Although we do not put {\it a priori} assumptions for 
spin and spatial configurations of
4 particles,
the $\vartheta^+(J^P=1^+)$ wave function obtained by the energy variation
is dominated by 
the component with the spin-zero $(ud)_{\bar{3}}$ and the spin-one
$(\bar{s}\bar{s})_3$ in the spatially symmetric orbit, $(0s)^4$.
The spin-zero $(ud)_{\bar{3}}$ gain the color-magnetic
interaction, while only the spin-one
configuration is allowed in the spatially symmetric 
$(\bar{s}\bar{s})_3$ pair. Therefore this is consistent 
with the naive expectation in the diquark picture.

We comment on the accuracy of approximations Eqs.
\ref{eq:3q},\ref{eq:4q} and \ref{eq:5q} for the tube length 
in the Hamiltonian. 
As discussed in a previous paper \cite{ENYO-penta}, 
the tube length ($L_f$) is reasonably simulated by the 
approximated tube length $L_{app}$ given by Eqs.
\ref{eq:3q},\ref{eq:4q} and \ref{eq:5q}, while 
$L_{app}$ is exactly equal the $L_f$ in the $q\bar{q}$ system.
If we assume the harmonic oscillator $(0s)^{N_q}$ 
configurations and ignore the antisymmetrization
of quarks, we can calculate the ratio of the $L_{app}$ 
to the exact tube length ($L_f$), which is denoted by 
$L_{app}^{(0s)}/L_f^{(0s)}$. 
The ratio $L_{app}^{(0s)}/L_f^{(0s)}$ is 0.91, 0.86 and 0.84 for the
$q^3$, $(qq)_{\bar{3}}(\bar{q}\bar{q})_3$ and 
$(qq)_{\bar{3}}(qq)_{\bar{3}}\bar{q}$ systems, respectively.
In order to examine the effect of this factor on the tetraquark mass, 
we scale the tube length as 
$L_f \approx L_f^{(0s)}/L_{app}^{(0s)}\times L_{app}$ and estimate the
expectation value of the Hamiltonian for the present wave functions. 
With the obtained energies, we retune the $M_0$ by 
fitting the $\rho$-meson, nucleon and pentaquark masses, and reexamine the
$\vartheta$ mass. We then find that the modification of the 
tetraquark mass by the scaled $L_f$ is slight;
the $\vartheta^+(J^P=1^+)$
mass can decrease by 10 MeV for case (i) and 4 Mev for case (ii).

\subsection{$f_1$-meson}

Here, we discuss the possibility of a $f_1$($J^PC=1^{++}$)-meson 
with the $4q$ component. Since its dominant decay mode should 
be $K\bar{K}^*$, it may have some analogy with the tetraquark 
$\vartheta(J^PC=1^{++})$.
If we ignore the $q\bar{q}$ annihilation, we can calculate the mass of the 
$J^{PC}$=$1^{++}$ $(us)_{\bar{3}}(\bar{u}\bar{s})_3$ state 
within the present framework 
in the same way as for the tetraquark $\vartheta$-meson.
The $(us)_{\bar{3}}(\bar{u}\bar{s})_3$($J^{PC}$=$1^{++}$) state corresponds
to an isoscalar $f_1$-meson and an isovector 
$a_1$-meson, which degenerate in the present Hamiltonian.
We concentrate on the $f_1$-meson in the present paper.
By using the same zero-point energies 
(i) $M^0_{[qq][\bar{q}\bar{q}]}=1785$ MeV and
(ii) $M^0_{[qq][\bar{q}\bar{q}]}=1679$ MeV,  
the $(us)_{\bar{3}}(\bar{u}\bar{s})_3$ state with 
$J^{PC}$=$1^{++}$ for the
$f_1$-meson is calculated to be 1.45 GeV in case(i) and 1.54 GeV in case(ii) 
(Fig.~ \ref{fig:4q}).

There are various theoretical
interpretations of scalar and axial-vector mesons as $P$-wave
$q\bar{q}$ states, $4q$ states and hybrid $q\bar{q}g$ states.
In the mass region 1$\sim$1.6 GeV, three $f_1$-mesons,
$f_1(1285)$, $f_1(1420)$, and $f_1(1510)$ are known,
though the $f_1(1510)$ is not well established \cite{PDG}.
In the $P$-wave $q\bar{q}$ state, two $f_1$-mesons are expected to 
appear in this energy region as partners in the $q\bar{q}$ nonet. 
It is considered that the lower one is dominated by 
the light-quark component ($n\bar{n}\equiv u\bar{u}+d\bar{d}$), 
and the major component of the higher one is the $s\bar{s}$ state.
In the standard interpretation, the lowest $f_1(1285)$ is 
regarded as the $n\bar{n}$ state.
On the other hand, $f_1(1420)$ and $f_1(1510)$ are candidates 
for the partner of the $f_1(1285)$ in the $q\bar{q}$ nonet, 
but the assignment is not yet confirmed. 

In the constituent quark model calculation of $q\bar{q}$ systems
\cite{Isgur-2q},
the masses of two $1^{++}$ states in the $P$-wave 
$q\bar{q}$ nonet are 1.24 and 1.48 GeV. 
The theoretical mass spectra of the $1^{++}$ $q\bar{q}$ states 
seems to be consistent with the experimental ones
if $f_1(1510)$ is assigned to be a partner of the $f_1(1285)$ in the 
flavor nonet.
This is consistent with the assignment in Ref.\cite{Godfrey}.
On the other hand, an alternative interpretation that the 
$f_1(1285)$ and $f_1(1420)$ are $q\bar{q}$
partners is claimed in Refs.\cite{close02,PDG,close97}.

These interpretations lead to an indication that one of 
$f_1(1420)$ and $f_1(1510)$ may be a non-$q\bar{q}$ meson, while the other 
can be understood as being partners of the $f_1(1285)$ 
in the conventional $P$-wave $q\bar{q}$ states.
In the present calculation of the $(us)_{\bar{3}}(\bar{u}\bar{s})_3$  state
with $J^{PC}$=$1^{++}$, the theoretical 
mass in case~ (i) 1.45 GeV seems to be consistent with the $f_1(1420)$, while
the mass in case~ (ii) 1.54 GeV energetically corresponds to the $f_1(1510)$,
as shown in Fig.~ \ref{fig:4q}. The present results suggest 
the assignment of a $f_1$-meson
in the 1.4$\sim$1.6 GeV mass 
region as the $4q$ state.

\subsection{Width of $\vartheta^+$-meson}
\label{subsec:width}

As mentioned above, we suggest that the $\vartheta^+(J^P=1^{+})$-meson
may appear in the energy region $\sim$1.4 GeV near the $KK^*$ threshold.
The expected decay modes are $KK^*$ and $KK\pi$.
The width for the $KK\pi$ decay can be enhanced by the broad resonance,
$\kappa(800)$ state, in the scalar $K\pi$ channel.
On the other hand, the phase space for the direct three-body decay 
is generally suppressed. In fact, the typical width of three-body 
decays is of the order of a several MeV in the $\omega$-meson 
width, for example.
Therefore, if the branchings into two-hadron decays, 
$KK^*$ and $K\kappa(800)$, 
are small enough, the width should be narrow.
In order to discuss the stability of the $\vartheta^+$-meson,
we, here, consider only the two-hadron decay modes.
We should note that the decay mechanism of  
the $\vartheta^+(J^P=1^{+})$ may be analogous with that of the 
$f_1$-meson with the $us\bar{u}\bar{s}(J^{PC}=1^{++})$ state,
where the $K\bar{K}^*$ and $K\bar{K}\pi$($K\bar{\kappa}$) 
are expected to be dominant decay modes.
In fact, in the decay of $f_1(1420)$ and $f_1(1510)$, 
which are the candidates of the 
$us\bar{u}\bar{s}(J^{PC}=1^{++})$, as mentioned before, 
the $K\bar{K}^*$ and/or $K\bar{K}\pi$ modes are 
experimentally seen, and the former is dominant in the $f_1(1420)$ 
decay modes\cite{PDG}.
For decay into the $K\bar{K}^*$, the $S$-wave channel is allowed, while 
the $K\bar{\kappa}$ decays should be a $P$-wave.

First, we give a rough estimation of the $\vartheta^+$ width for 
the $KK^*$ decay by assuming that the coupling
for $f_1\rightarrow K\bar{K}^*$ ($g_{f_1 K\bar{K}^*}$)  
is the same as that for $\vartheta^+ \rightarrow KK^*$ 
($g_{\vartheta KK^*}$).
The width is approximated by the product of the 
coupling and the phase space. 
We take into account only the $S$-wave decay, and estimate the phase
space for the $KK^*$ decay by the imaginary part of the
one-loop self-energy integral $I(p)$ for the scalar mesons,
\begin{equation}
{\rm Im}[I(p)] =\frac{1}{16\pi^2} {\rm Im}\left[-a_1 
\ln(m^2_1)-a_2\ln(m^2_{2})
-q\ln\frac{(q+a_1)(q+a_2)}{(q-a_1)(q-a_2)} \right], 
\end{equation}
where,
\begin{eqnarray}
& q\equiv\frac{1}{2}\left[\left(1-\frac{(m_1+m_{2})^2}{p^2}\right)
\left(1-\frac{(m_1-m_{2})^2}{p^2}\right)\right]^{\frac{1}{2}}\nonumber\\
& a_1\equiv\frac{1}{2}(1-\frac{m_1^2-m_{2}^2}{p^2})\nonumber\\
& a_2\equiv\frac{1}{2}(1-\frac{m_{2}^2-m_{1}^2}{p^2}), \label{eq:Im}
\end{eqnarray}
and $m_1$ and $m_2$ are the masses of the daughter particles.
We take the rest frame $p=(p_0,{\bf 0})$.
As is well known, if the $m_1$ and $m_2$ are real values, Im[$I(p)$] is
equal to the usual phase space $q$ above the threshold ($p_0>m_1+m_2$),
and is zero below the threshold
($|m_2-m_1|<p_0<m_1+m_2$).
However, since the daughter particle $K^*$ has a width of 
$\Gamma_{K^*}\sim 50$ MeV for $K\pi$ decays, 
the phase space should be finite, even below the threshold.
In order to estimate the effect of $\Gamma_{K^*}$ on the phase space,
we take the masses of the daughter particles 
to be $m_1=m_{K^*}-i\Gamma_{K^*}/2$ and $m_2=m_K$ in Eq.~ \ref{eq:Im},
as is done in Ref.\cite{Hidaka}. 
We use the masses of $K$ and $K^*$ as $m_K=0.495$ GeV and $m_{K^*}=0.892$ GeV,
and evaluate the width of the parent particle as
$\Gamma=2g^2{\rm Im}[I(m_0)]$, where $m_0$ is the parent particle mass and
$g$ is the coupling. 

\begin{figure}
\noindent
\centerline{\epsfxsize=0.4\textwidth\epsffile{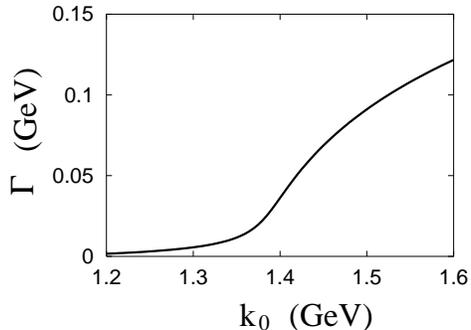}}
\caption{\label{fig:Im}
Energy dependence of the two-meson $KK^*(892)$ decay 
width $\Gamma$. The coupling is chosen to fit the
full-width $\Gamma=55$ MeV of the $f_1(1420)$ at $m_0=1.42$ GeV,
and is assumed to be energy independent.}
\end{figure}

We show the 
mass $m_0$ dependence of the width $\Gamma(m_0)$ 
for the $KK^*$ decays in Fig.~ \ref{fig:Im}, where 
the coupling $g$ is adjusted to reproduce 
the experimental full width $\Gamma_{f_1(1420)}=56$ MeV at $m_0=1.42$ GeV,
and is assumed to be energy independent.
If we adopt the case~ (i) calculation and the assignment of the $f_1(1420)$
as the $4q$ state, the width of the $\vartheta^+(1^{+})$ is estimated to be
$\Gamma_{\vartheta}\sim 20$ MeV.
When we assign the $f_1(1510)$ as the $4q$ state 
and choose the coupling constant $g$ 
to fit the width 100 MeV of the $f_1(1510)$ at $m_0=1.51$ GeV, 
we obtain $\Gamma_{\vartheta}\sim 80$ MeV
based on the case~ (ii) calculation. Here, we adopt the upper limit of the
$f_1(1510)$ width in Ref.\cite{PDG}, though there still remains an ambiguity
in the experimental data.

Next, we discuss the effect of $K\kappa$ decay on the width.
In the above estimation of $\Gamma_{\vartheta}$, 
the effect of this mode is already included
in the total width of the $f_1$-meson. 
In contrast to the $S$-wave decay in the $KK^*$ channel, the $1^+$ states
decay into the $P$-wave $K\kappa$ state. Even though the $P$-wave decay 
is unfavored compared to the $S$-wave decay, we should consider its effect,
because the $K\kappa$ threshold ($\sim 1.3$ GeV) is lower than 
the $KK^*$ threshold (1.39 GeV). 
We again assume that the coupling
for $f_1\rightarrow K\bar{\kappa}$ is the same as 
that for $\vartheta^+ \rightarrow K\kappa$, and estimate the width 
by using the ratio of 
the phase space for the $\vartheta$($m_0=1.37$ GeV), $q_{\vartheta}$,
to that for the $f_1(1420)$($m_0=1.42$ GeV), $q_{f_1}$, as
$\Gamma_\vartheta=\Gamma_{f_1}\times q_{\vartheta}/q_{f_1}$. 
Since $\kappa(800)$ is a broad resonance with $\Gamma=300\sim 800$ MeV, 
we may not apply the previous method by the 
single-pole approximation of the propagator 
in the one-loop diagram to estimate the phase space.
Instead, we consider the phase space for the $P$-wave decay,
$|m_0-m_\kappa-m_K|^{3/2}$, and take into account 
the broad mass range of $m_\kappa$, where the $m_\kappa$ is 
the $\kappa$ mass and is a real value.
When we adopt the case (i) calculation,
we obtain the ratio ($q_{\vartheta}/q_{f_1}$) of 
the phase space for the $\vartheta$($m_0=1.37$ GeV) 
to that for the $f_1(1420)$($m_0=1.42$ GeV) as $0.5$ for 
$m_\kappa=800$ MeV. The ratio $q_{\vartheta}/q_{f_1}$
is $0.7$ for the lower limit $m_\kappa=640$ ($K\pi$ threshold) MeV, while
$q_{\vartheta}/q_{f_1}=0$ for $m_\kappa\ge 900$ MeV.
On the avarage, the ratio $q_{\vartheta}/q_{f_1}$ in the $K\kappa$ decay 
is almost the same as that in the $KK^*$ decay. Even if the branching ratio
of the $K\kappa$($K\bar{\kappa}$) mode is 100\% in  
$\vartheta^+(1^{+})$($f_1$), an upper limit of $\Gamma_{\vartheta}= 40$ MeV 
is obtained from the $\Gamma_{f1}=56$ MeV and the lower limit of $m_\kappa$.
Also, in calculation (ii), 
the phase space ratio ($q_{\vartheta}/q_{f_1}$) for  
$f_1(1510)\rightarrow K\bar{\kappa}$ to that for the 
$\vartheta \rightarrow K\bar{\kappa}$ in the mass range $m_\kappa \ge 640$ MeV 
is calculated as $q_{\vartheta}/q_{f_1}<0.8$, which is
consistent with that in the 
$KK^*$ decay. As a result, it is concluded that the $\vartheta$-meson
width is estimated to be ${\cal O}(20-80 MeV)$, no matter how great are the 
branching ratio in the $KK^*$ channel and that in the $K\kappa$ channel. 

In the above discussion, 
we roughly evaluate the width of the $\vartheta^+(1^{+})$-meson
by assuming the same coupling for $\vartheta^+(1^{+})\rightarrow KK^*$ 
and $\vartheta^+(1^{+})\rightarrow K\kappa$ 
as those for $f_1\rightarrow K\bar{K}^*$ and
$f_1\rightarrow K\bar{\kappa}$, respectively.
In the present estimation we choose the couplings 
to fit the full width of $f_1$. In the case that 
the couplings are enhanced 
via the annihilation and creation of a $q\bar{q}$ pair in the $f_1$-meson, 
the width of the $\vartheta^+(1^{+})$-meson might be smaller than the
present estimation.

\section{Summary and discussion}

We discussed the possibility of the $J^P=1^+$ state of the
isoscalar tetraquark (S=+2), $\vartheta^+$-meson, 
with the $ud\bar{s}\bar{s}$ content.
If the pentaquark $\Theta^+(1540)$ has the 
$(qq)_{\bar{3}}(qq)_{\bar{3}}\bar{q}$ configuration,
the $\vartheta^+(J^P=1^+)$
is expected to exist at an energy lower than, or close to, 
the pentaquark $\Theta^+(1540)$ mass. 
This leads to a possible appearance of the $\vartheta^+$-meson as a 
resonant state, which may be observed 
in the $K^+K^+\pi^-$ channel. 

We investigated the $\vartheta^+(J^P=1^+)$
with a constituent quark model.
The flux-tube quark model with antisymmetrized molecular
dynamics(AMD) was applied to $4q$ systems in the same way as in 
the pentaquark study in Ref.\cite{ENYO-penta}.
Based on the picture of a flux-tube model, 
we solved the $4q$ dynamics in the $(qq)_{\bar{3}}(\bar{q}\bar{q})_3$ model 
space by the variational method.
The results suggest that the $\vartheta^+(J^P=1^+)$ may 
exist at around 1.4 GeV. 
Since the predicted mass of the $\vartheta^+(J^P=1^+)$ 
is close to the lowest ($KK^*$) threshold in the 
allowed two-hadron decays, the present results imply that the
$\vartheta^+(J^P=1^+)$ may exist as a resonance, and its width may 
be not very broad.

We also calculated the $J^{PC}=1^{++}$ $(us)_{\bar{3}}(\bar{u}\bar{s})_{3}$ 
state, which is associated with a non-$q\bar{q}$ $f_1$-meson.
The calculated mass of the $f_1$ with the $4q$ configuration
suggests an interpretation that the $4q$ state may correspond to 
one of the $f_1$-mesons in the $1.4\sim 1.6$ mass region.
$f_1(1420)$ and $f_1(1510)$ are candidates of the $4q$ state. 
If we assume that
the couplings for $f_1\rightarrow K\bar{K}^*$ and $f_1\rightarrow 
K\bar{\kappa}^*$ are the same as those for 
$\vartheta^+(J^P=1^+)\rightarrow KK^*$ and 
$\vartheta^+(J^P=1^+)\rightarrow K\kappa$, respectively, 
we can evaluate the width of the $\vartheta^+$
to be 20-80 MeV from the phase space for these two-hadron decay modes.
Provided that the coupling for the
two-hadron decays in the $\vartheta^+(J^P=1^+)$ is small enough,
the dominant decay should be a direct three-hadron decay, 
$\vartheta^+\rightarrow KK\pi$,
with a small phase space, and hence the $\vartheta^+(J^P=1^+)$
should have a small width.

Recently, the $\vartheta^+$-meson was discussed by Burns et al.\cite{Close-4q}
and by Karliner and Lipkin \cite{karliner04}.
In Ref.\cite{karliner04}, it was mentioned 
that the $\vartheta(J^P=0^+)$ can be narrow, because
the lowest allowed decay mode is a four-body $KK\pi\pi$ channel with a 
small phase space. However, the $J^P=0^+$ state is forbidden
in the isoscalar $ud\bar{s}\bar{s}$ system within 
spatially symmetric configurations, and hence, 
the $\vartheta(J^P=0^+)$ is expected to be energetically unfavored.
Burns et al. predicted the $\vartheta^+(J^{P}=1^{-})$ with $L=1$ 
at $\sim$ 1.6 GeV with a
width of $O(10-100)$ MeV, which can decay into $K^+K^0$.
Although the color-magnetic attraction may be larger 
in the $\vartheta(J^P=1^-)$ state than in the $\vartheta(J^P=1^+)$ state, 
the $\vartheta(J^P=1^-)$ must have the $L=1$ excitation energy.
Another claim for the $\vartheta^+(J^P=1^-)$ state is that it 
can decay into $P$-wave $KK$ states. The centrifugal barrier
may not be high enough to stabilize the state much above the threshold energy. 
Therefore, we consider that the 
$\vartheta^+(J^P=1^+)$ is a better candidate of
narrow tetraquarks.

Our calculations of the tetraquark $\vartheta^+(J^{P}=1^{+})$ and the 
pentaquark $\Theta^+$ are based on the color-configurations 
$(qq)_{\bar{3}}(\bar{q}\bar{q})_3$ and 
$(qq)_{\bar{3}}(qq)_{\bar{3}}\bar{q}$. We ignore $(qq)_6$ configurations 
because the $(qq)_6$ is an excited configuration, and is 
unfavored in the strong-coupling picture.
We should remark that the preference of the spin-parity $J^{P}=1^{+}$ 
in the $\vartheta^+$ system does not depend on the color configuration. 
This is because, in the isoscalar $ud\bar{s}\bar{s}$ with
a spatially symmetric configuration,
other spin parities are forbidden, and therefore 
the spin parity is uniquely determined to be $J^P=1^+$.
This means that, even in the color-configuration 
$(ud)_{6}(\bar{s}\bar{s})_{\bar{6}}$, as suggested in 
Refs.\cite{Close-4q,karliner04}, only $J^P=1^+$ is allowed 
in the spatially symmetric $\vartheta$ state.
If the $\Theta^+$ has a triquark-diquark structure 
$((ud)_{6}\bar{s})_{3}(ud)_{\bar{3}}$ with the relative $P$-wave motion 
as proposed by Karliner and Lipkin\cite{karliner}, it is expected that
the $\vartheta(J^P=1^+)$ mass with a spatially symmetric 
$(ud)_{6}(\bar{s}\bar{s})_{\bar{6}}$ configuration would be close to, or 
could be smaller than, the $\Theta^+$ mass due to the kinematic energy gain.
Another important channel is the meson-meson $(q\bar{q})_{1}(q\bar{q})_1$
configuration, though it can be expressed by a linear combination of 
the $(ud)_{\bar{3}}(\bar{s}\bar{s})_3$ and 
$(ud)_{6}(\bar{s}\bar{s})_{\bar{6}}$ configurations. 
If the $\vartheta^+(J^{P}=1^{+})$ mass is smaller than the
lowest meson-meson threshold $KK^*$, mixing of the 
$(q\bar{q})_{1}(q\bar{q})_1$ state may not have a significant effect on the
$\vartheta$ width, because 
the $\vartheta^+(J^{P}=1^{+})$ is bound in the $KK^*$ channel.
In the case that the $\vartheta^+(J^{P}=1^{+})$ is heavier than the 
$KK^*$ threshold, channel mixing should be taken into consideration
to estimate its width.
In \ref{subsec:width}, we gave an analysis of the widths by assuming
that the decay mechanism and the coupling
in the $\vartheta^+(J^P=1^{+})$ are the same as those in
the $f_1$-meson. Within this assumption, the channel-coupling 
effect on the widths is effectively included in the adopted total width 
of the $f_1$-meson.
For a further detailed discussion of the stability of 
the $\vartheta(J^P=1^+)$, coupled calculations of 
different color configurations are required.

We should point out that 
the allowed decay channels are different among these
three predictions ($J^P=1^+$, $1^-$, and $0^+$ states) of 
the $\vartheta^+$-mesons. 
For the $\vartheta^+(J^P=1^+)$, $\vartheta^+(J^P=1^-)$, and 
$\vartheta^+(J^P=0^+)$,
the decay modes are $KK\pi$, $KK$ and $KK\pi\pi$, respectively.
In order to establish the $ud\bar{s}\bar{s}$ content
in the tetraquark $\vartheta^+$-meson,
it is necessary to observe not $K^0$, but 
at least two $K^+$s, because the $K^0$ contains the $s\bar{d}$ component 
as well as the $d\bar{s}$. 
In that sense, the $K^+K^+\pi^-$ decay from 
the $\vartheta$($J^{P}=1^{+}$) predicted in the present work
is suitable for an experimental tetraquark search.

In the constituent quark model calculations of Refs.\cite{Isgur-4q,carlsonb}, 
there is no indication for the existence of multiquark hadrons, 
except for the $f_0(980)$ and $a_0(980)$ as $K\bar{K}$ molecules
\cite{Isgur-4q}, while interpretations of scalar mesons, 
like $f_0(600)$, $\kappa(800)$ and $D_{sJ}$(2317), by 
four-quark states have been suggested 
in several quark model calculations
\cite{Jaffe-4q,Jaffe-h,Jaffe79,beveren86,beveren03,Maiani,terasaki}.
In Ref.\cite{carlsonb}, the absence of the pentaquark
was claimed, which is inconsistent with the observation of the pentaquark
$\Theta^+$.  
In order to explain the absolute mass of the 
$\Theta^+(1540)$ within constituent quark models, we need an extra attraction 
for the multiquark system in addition to a pair-wise interaction. 
In the present calculation, we did not obtain a quantitative 
value of this extra attraction from fundamental theory; instead,
the many-body potential was taken into account 
in terms of the flux-tube potential, and we phenomenologically
evaluated it by using the observed $\Theta^+$ mass as a input. 

In the present work, we used a simple Hamiltonian with 
the confining force, and the Coulomb and
color-magnetic terms in the OGE potential.
For a systematic description of the hadron spectra,
there still remain such problems as fine tuning of
the interaction parameters and inclusion of the tensor and spin-orbit interactions
in the Hamiltonian.
To obtain theoretical insights into multiquark hadron physics, 
an experimental search for the tetraquark 
is necessary as well as further experimental studies 
on the pentaquark.
We conclude that the $\vartheta^+(J^P=1^+)$-meson
is proposed as a good candidate of the tetraquark, 
which would be observed in the $K^+K^+\pi^-$ decay channel.

\acknowledgments

The authors would like to thank to H. Nemura and H. Hidaka 
for valuable discussions.
This work was supported by Japan Society for the Promotion of 
Science and Grants-in-Aid for Scientific Research of the Japan
Ministry of Education, Science Sports, Culture, and Technology.


\def\Ref#1{[\ref{#1}]}
\def\Refs#1#2{[\ref{#1},\ref{#2}]}
\def\npb#1#2#3{{Nucl. Phys.\,}{\bf B{#1}},\,#2\,(#3)}
\def\npa#1#2#3{{Nucl. Phys.\,}{\bf A{#1}},\,#2\,(#3)}
\def\np#1#2#3{{Nucl. Phys.\,}{\bf{#1}},\,#2\,(#3)}
\def\plb#1#2#3{{Phys. Lett.\,}{\bf B{#1}},\,#2\,(#3)}
\def\prl#1#2#3{{Phys. Rev. Lett.\,}{\bf{#1}},\,#2\,(#3)}
\def\prd#1#2#3{{Phys. Rev.\,}{\bf D{#1}},\,#2\,(#3)}
\def\prc#1#2#3{{Phys. Rev.\,}{\bf C{#1}},\,#2\,(#3)}
\def\prb#1#2#3{{Phys. Rev.\,}{\bf B{#1}},\,#2\,(#3)}
\def\pr#1#2#3{{Phys. Rev.\,}{\bf{#1}},\,#2\,(#3)}
\def\ap#1#2#3{{Ann. Phys.\,}{\bf{#1}},\,#2\,(#3)}
\def\prep#1#2#3{{Phys. Reports\,}{\bf{#1}},\,#2\,(#3)}
\def\rmp#1#2#3{{Rev. Mod. Phys.\,}{\bf{#1}},\,#2\,(#3)}
\def\cmp#1#2#3{{Comm. Math. Phys.\,}{\bf{#1}},\,#2\,(#3)}
\def\ptp#1#2#3{{Prog. Theor. Phys.\,}{\bf{#1}},\,#2\,(#3)}
\def\ib#1#2#3{{\it ibid.\,}{\bf{#1}},\,#2\,(#3)}
\def\zsc#1#2#3{{Z. Phys. \,}{\bf C{#1}},\,#2\,(#3)}
\def\zsa#1#2#3{{Z. Phys. \,}{\bf A{#1}},\,#2\,(#3)}
\def\intj#1#2#3{{Int. J. Mod. Phys.\,}{\bf A{#1}},\,#2\,(#3)}
\def\sjnp#1#2#3{{Sov. J. Nucl. Phys.\,}{\bf #1},\,#2\,(#3)}
\def\pan#1#2#3{{Phys. Atom. Nucl.\,}{\bf #1},\,#2\,(#3)}
\def\app#1#2#3{{Acta. Phys. Pol.\,}{\bf #1},\,#2\,(#3)}
\def\jmp#1#2#3{{J. Math. Phys.\,}{\bf {#1}},\,#2\,(#3)}
\def\cp#1#2#3{{Coll. Phen.\,}{\bf {#1}},\,#2\,(#3)}
\def\epjc#1#2#3{{Eur. Phys. J.\,}{\bf C{#1}},\,#2\,(#3)}
\def\mpla#1#2#3{{Mod. Phys. Lett.\,}{\bf A{#1}},\,#2\,(#3)}
\def\etal{{\it et al.}}

\end{document}